\documentclass[aps,twocolumn]{revtex4}

\usepackage{dcolumn}
\usepackage{bm}
\usepackage{graphicx}
\newcommand{\nc}{\newcommand}
\nc{\bc}{\begin{center}}
\nc{\ec}{\end{center}}
\begin{document}
\title{Collective Two-Atom Effects and Trapping States in the Micromaser}
\author{Per K. Rekdal $^{a,}$}
  \email{per.rekdal@uni-graz.at}
\author  {Bo-Sture K. Skagerstam$^{b,}$}
  \email{boskag@phys.ntnu.no}

\affiliation
  {\vspace{1mm} \small $^a$ Institut f\"ur Theoretische Physik,
                            Karl-Franzens-Universit\"at Graz,
                            Universit\"atsplatz 5, A-8010 Graz, Austria
                       \\  
                       \\
                \small $^b$ Complex Systems and Soft Materials
  Research Group, Department of Physics,
                            The Norwegian University of Science and Technology, 
                            N-7491  Trondheim, Norway
  }

{\begin{abstract}
     \small 
      \noindent 
          We investigate signals of trapping states in the micromaser system
     in terms of the average number  $\langle n \rangle$ of cavity photons
     as well
     as a suitably defined correlation length of atoms leaving the cavity. In the
     description of collective two-atom effects we allow the mean number, $\epsilon$,
     of pump atoms inside the cavity during the characteristic atomic cavity transit time  to be as large as of  order
     one. 
     The master equation we consider, which describes the micromaser including collective 
two-atom effects, still exhibits trapping states for
     $\epsilon \neq 0$, even for a mean 
     number of atoms inside the cavity close to one. We, however, argue  more importantly that the trapping states are 
     more pronounced in terms of the correlation length as compared to $\langle n \rangle$, i.e. we suggest that trapping states can be more clearly revealed 
     experimentally in terms of the atom correlation length. For axion detection in the micromaser this observable may therefore be an essential ingredient.
     \\[5mm]
\noindent PACS numbers 32.80.-t, 42.50.-p, 42.50.Ct
     $~~~~~~~~~~~~~~~~~~~~~~~~~~~~~~~~~~$ 
\end{abstract}
}

\maketitle

\bc{
\section{INTRODUCTION}
\label{sec:introd}}\ec
The micromaser \cite{meschede85,Filipowicz86,raithel94} belongs to one of the simplest 
    systems in cavity quantum electrodynamics. For this reason  plenty
    of  experimental as well as theoretical investigations have been carried out. 
    A micromaser consists of a beam of identical
    two-level atoms passing through a cavity and interacting with a single mode of the
    electromagnetic field. Usually the atomic flux is so small that at most one atom interacts 
    with the cavity mode simultaneously. 
    This is the situation in the standard one-atom micromaser theory.
    For larger values of the atomic flux,  and also due to statistical 
    distribution of atomic velocities of
    the pump atoms, in an actual experimental situation there is,
    however, a finite probability for 
    two or more atoms to be simultaneously present in the cavity.
    An extension of the standard micromaser theory to include collective 
    effects of many atoms is therefore needed. 
    This has e.g. been done in 
Refs.\cite{wehner94,haake97,casagrande99a,casagrande99b,schieve_03,ulzega_03,skagerstam_06},
    where two-atom effects have been taken into account.
    Investigations of collective effects have been simulated by their significance on the
    mean photon number in the cavity field. 
    At zero temperature, so called trapping states of the cavity radiation
    field \cite{Filipowicz86,slosser_90} may appear in the micromaser system \cite{W&V&H&W,W&V&H&W_2}. These trapping states
    corresponds to a truncation of the steady-state photon distribution, and
    thereby to  the appearance of a minimum  in the mean photon number as a function of the atom cavity transit time. The preparation of such trapping states, in particular the vacuum trapping state, is essential in the design of axion detectors using the micromaser system (see Ref.\cite{jones_2006} and references cited therein).
    It has been argued that the appearance of trapping states maybe
    suppressed when including collective atom effects \cite{haake97,schieve_03,skagerstam_06}.
    Recently, there has, however, been some disagreement on the degree
    of this suppression \cite{schieve_03,ulzega_03}.

    Several effects in the micromaser can affect the system dynamics and destabilize the trapping
    states by increasing the number of the cavity photons. 
    The trapping effect in terms of the mean photon number is
    therefore difficult to 
observe experimentally.
    In the present paper we will therefore not only consider the photon number but also  a suitably defined 
    correlation length of the atoms leaving the micromaser.

    The paper is organized as follows. In Section \ref{sec:dynamics} we outline the dynamics of the standard
    one-atom micromaser, i.e. we present the well known master equation for the photon field in this case.
    In the following Section we extend this theory to the two-atom situation, with a finite probability of having
    two atoms simultaneously in the cavity. The numerical results are presented in Section \ref{sec:results} 
    together with a discussion.

\vspace{1cm}
\bc{
\section{ONE-ATOM MICROMASER DYNAMICS}
\label{sec:dynamics}}\ec

   The continuous time formulation of the dynamics of the one-atom micromaser is very well known (see e.g. Refs.\cite{	Lugiato87,ElmforsLS95}).
We now outline a typical realization of the one-atom micromaser system. The pump atoms which
   enter the cavity are at resonance with the radiation field of the cavity.
   The vector $p$ formed by the diagonal density matrix elements of 
   the photon field then obeys the master equation

\begin{equation} \label{master_eq} 
   \frac{dp}{dt} = - \gamma {\cal L}_1 p~~,   
\end{equation}

    \noindent  where

\begin{equation} \label{L_1}
   {\cal L}_{1}  \equiv  L_C  -  N \, ( \, {\cal U}_1 - 1  \, ) ~ .
\end{equation}

    \noindent 
    Here the dimensionless atomic flux parameter is $N=R/\gamma$, where $R$ is the rate of
    injected atoms and $\gamma$ is the damping rate of the cavity. The
    cavity damping is described by the matrix

\begin{eqnarray}
  (L_C)_{nm} &=& (n_b+1)[ \, n \delta_{n,m} - (n+1) \delta_{n+1,m} \, ] 
  \\ \nonumber
  &+& n_b[ \, (n+1)\delta_{n,m} - n \delta_{n,m+1} \, ] ~ ,
\end{eqnarray}

   \noindent
   where the one-atom operator  matrix ${\cal U}_1$ is given by

\begin{eqnarray}     \label{1_atom_op}
  ({\cal U}_1)_{nm} &=& (1-q_{n+1}) \delta_{n,m} + q_{n} \delta_{n,m+1}   ~ ,
\end{eqnarray}

   \noindent
   provided that all the pump atoms are prepared in the excited state. Here we have defined 
   $q_n = \sin^2( \, g\tau \, \sqrt{n}) = \sin^2( \, \theta \, \sqrt{n/N} \, )$ and introduced the natural dimensionless pump parameter 
   $\theta \equiv g\tau \sqrt{N}$ in terms of the atomic transit time $\tau$. Furthermore, $g$ is the single 
   photon Rabi frequency at zero detuning of the Jaynes-Cummings-model \cite{Jaynes63}. Due to conservation of 	probability, the matrices above satisfy

\begin{eqnarray} \label{sums1_eq}
  \sum_{n=0}^{\infty} ~ (L_{C})_{nm} &=& 0 ~ ,
\end{eqnarray}
and
\begin{eqnarray} \label{sums2_eq}
  \sum_{n=0}^{\infty} ~ ({\cal U}_1)_{nm} &=& 1 ~,
\end{eqnarray}

   \noindent
   for any given integer $m \geq 0$,  as can be verified
   analytically. The stationary solution of Eq.(\ref{master_eq})
   is then given by \cite{Filipowicz86,Lugiato87}

\begin{equation} \label{p_n_eksakt}
  {\bar p}_n = {\bar p}_0 \prod_{m=1}^{n}  \frac{n_b \, m +  N q_m}{(1+n_b) \, m } ~ . 
\end{equation}

   \noindent The overall constant ${\bar p}_0$ is determined by $\sum_{n=0}^{\infty}{\bar p}_n =1$.
   The mean number of photons in the cavity is given by  
   $\langle n \rangle = \sum_{n=0}^{\infty}{\bar p}_n n $.

\vspace{1cm}
\bc{
\section{TWO-ATOM MICROMASER DYNAMICS}
\label{sec:collective_2}}
\ec

    Since the average number of pump atoms inside the cavity is $\epsilon \equiv \tau R =\gamma\sqrt{N}\theta/g$,
    this parameter naturally parameterize the probability of collective pump atom effects. 
    In the present paper, we are not restricted to $\epsilon \ll 1$ as in Refs.\cite{haake97,skagerstam_06}, and we will even consider a case with
    $\epsilon \approx 1$.

    Let us assume that the atoms arrive at the micromaser cavity with Poissonian arrival statistics.   
    The probability that the distance between two atoms exceeds the
    length of the micromaser cavity is 
    then $\exp{(-\epsilon)}$ (see e.g. Ref.\cite{wehner94}). A one-atom
    event occurs whenever an atom is separated by at least the cavity length
    from both its neighbours in the beam. Likewise, a two-atom event
    occurs whenever two atoms are separated from each
    other by less than the cavity length and the nearest neighbours of
    the pair are further away then the cavity length.
    The probability that a randomly-chosen atom in the beam is a $n$-atom event then is \cite{wehner94}

\begin{equation} \label{p_arr}
   P_n = e^{-2 \, \epsilon} \, [ \, 1 - e^{-\epsilon} \, ]^{n-1}  ~ .
\end{equation}

    \noindent
    The probabilities in Eq.(\ref{p_arr}) are normalized in such a way that the total atomic flux $R$ is fixed, i.e. $\sum_{n=1}^{\infty}\, n RP_n = R$.
    A generalization of the one-atom  master equation to one that includes one-atom events as well as
    two-atom events has then been proposed in Ref.\cite{schieve_03}, i.e.

\begin{equation} \label{master_2_atom}
   \frac{d p }{d t} = - \gamma \, {\cal L}_{2} \, p ~ ,
\end{equation}

   \noindent
   where

\begin{equation} \label{eq:cns}
   {\cal L}_{2}  \equiv  L_C  -  N \, \tilde{P}_1  \, ( \, {\cal U}_1
   - 1  \, )   -   
N \, \tilde{P}_2 \, ( \, {\cal U}_2 - 1  \, ) ~ .
\end{equation}

   \noindent
  Here the modified one- and two-atom probabilities, $\tilde{P}_1$ and $\tilde{P}_2$, are given by \cite{schieve_03}

\begin{eqnarray}
   \label{P_1_t}
   \tilde{P}_1  &=&  \frac{P_1}{P_1 + 2 P_2} ~ ,
   \\
   \label{P_2_t}
   \tilde{P}_2  &=&  \frac{P_2}{P_1 + 2 P_2} ~ ,
\end{eqnarray}
 \noindent  normalized in such a way that $ \tilde{P}_1+ 2\tilde{P}_2=1$.
   The one-atom operator ${\cal U}_1$ is given by Eq.(\ref{1_atom_op}).
   In particular, for $\tilde{P}_1 = 1$ and $\tilde{P}_2 = 0$ the two-atom generator ${\cal L}_2$ is reduced 
   to the one-atom generator as defined in Eq.(\ref{L_1}). 
   The two-atom operator ${\cal U}_2$ is defined in such a way that 
   ${\cal U}_2 \, \rho_f(t)$ gives the reduced density operator for the radiation field for a  two-atom event \cite{schieve_03}, i.e.

\begin{eqnarray}   \label{U_2_def}
   {\cal U}_2 \, \rho_f(t)  &\equiv&  \int_0^{\tau} ds ~ w(s) ~ 
                            \\  \nonumber
                            &\times& \mbox{Tr}_{\mbox{at}_1
                            \mbox{at}_2}  
\{ \, u_1(\mbox{at}_2,s) \, u_2(\tau-s) \, u_1(\mbox{at}_1,s)
                            \\  \nonumber
                            &\times& \rho_{\mbox{at}_1} \otimes
                            \rho_{\mbox{at}_2}  \otimes \rho_{f}(t) ~ \} ~ .
\end{eqnarray}

   \noindent
 Here $u_1(\mbox{at}_i,s)$, for $i=1,2$, denotes the combined one-atom ($\mbox{at}_i$) and radiation field unitary time-evolution operator for a time-interval $s$.
   Similarly, $u_2(\tau-s)$ denotes the combined two-atom- and radiation field unitary time-evolution operator for the time-interval $\tau - s$.
   Tracing over the atomic states in Eq.(\ref{U_2_def}) gives the reduced density operator of the radiation field for a two-atom event.
   The integrand in Eq.(\ref{U_2_def}) describes a two-atom event, in which case the two atoms arrive a time $s$ apart and overlap 
   for a time $\tau-s$ during which both atoms are present in the cavity. 
   Since we assume that the arrivals of the atoms in the cavity are uncorrelated events of a 
   Poissonian process, the two-atom events enters in an averaged form with the 
  conditional probability density distribution \cite{schieve_03}

\begin{eqnarray}    \label{f_p}
  w(s) =  R ~ \frac{e^{- R\, s}}{1-e^{-R\tau}} ~ ,
\end{eqnarray} 
i.e. the Poisson probability for two atoms arriving a time $s$ apart conditioned by an overlap during the atomic transit time $\tau$. The distribution $w(s)$ is therefore normalized to one in the atomic transit time $\tau$. 

In the quantum trajectory approach to collective two-atom effects \cite{casagrande99a,casagrande99b}, an ensemble average is calculated, where the overlap $s$ is randomly selected according to the probability distribution Eq.(\ref{f_p}). Provided the system is ergodic, such an ensemble average should correspond to a time-average according to Eq.(\ref{U_2_def}). In Refs.\cite{haake97,skagerstam_06} a similar time-average has been considered. Recently it has been debated  to what extent the micro-maser system is ergodic when multi-atom effects are taken into account \cite{schieve_03,ulzega_03}, at least when considering the average number of cavity photons number $\langle n \rangle$. Trapping states appear to be more suppressed in the quantum trajectory approach \cite{ulzega_03} as compared to the time-average approach \cite{schieve_03} as defined by Eq.(\ref{U_2_def}).  Below we will, however, consider an alternative observable, which is more clearly sensitive to the presence of trapping states than $\langle n \rangle$ but which, at the same time, is not very sensitive to small changes in the physical parameters of the micromaser system. For this reason we conjecture that  calculations of this observable in the quantum trajectory approach and the time-average approach will lead to similar results. In the present paper we will restrict ourselves to the conventional time-average approach as in Eq.(\ref{U_2_def}).

   Explicit and correct expressions for the evolution operators $u_1$ and $u_2$ can be found using 
   Ref.\cite{haake97}. A straightforward calculation then gives

\begin{eqnarray}
    ({\cal U}_2)_{nm}  &=&  \int_0^{\tau} ds ~ w(s) ~ ({\cal U}_2 (s))_{nm}  ~ ,
\end{eqnarray}

   \noindent
   where we make use of the notation

\begin{eqnarray}
  ({\cal U}_2(s))_{nm}   &=&     {\cal U}_{aa , n} \, \delta_{n,m} 
                         \\ \nonumber
                         &+&   {\cal U}_{ab ,n-1} \, \delta_{n-1,m}  +
                         {\cal U}_{bb , n-2}  \, \delta_{n-2,m} ~ .
\end{eqnarray}

   \noindent
In the expression for the  matrix element $({\cal U}_2(s))_{nm}$  we have defined

\begin{eqnarray} \label{u2_1}
        {\cal U}_{aa , n} &=& 
        \big| ~     [ 1 - 2 \epsilon_n s_n^2 ] \, \tilde{c}_{n+1}^2 + s_n^2 \tilde{s}_{n+1}^2
                  \nonumber \\
                && ~  - 
                2 \, \sqrt{ 2 \epsilon_n } \, c_n s_n \, \tilde{s}_{n+1} \, \tilde{c}_{n+1} ~ \big| ^2 ~ ,
\end{eqnarray}
and 
\begin{eqnarray} \label{u2_2}
        {\cal U}_{ab ,n} &=& 
        \big| ~    \tilde{s}_{n+1} \, \tilde{c}_{n+1} \, [ \, ( \, 1 +  2 \epsilon_{n} \, ) \, s_{n}^2 \, - 1 \, ] 
        \nonumber \\
        &&  ~  +  
        \sqrt{2 \epsilon_{n} } \, c_{n} s_{n} \, [ \, \tilde{s}_{n+1}^2 \, - \, \tilde{c}_{n+1}^2 \, ] ~ \big| ^2
         \nonumber \\ 
         &+&    
         \left|  -   [ \, c_{n}^2 \tilde{s}_{n+1} +   
         \sqrt{2 \epsilon_{n}}   c_{n-1} s_{n}  \tilde{c}_{n+1}  \, ] \, \tilde{c}_{n+2}^{\mbox{}} \right.
         \nonumber \\    
         && ~ + 
          [ ~  2 \sqrt{ \epsilon_{n} (1-\epsilon_{n}) } \, s_{n}^2 \, \tilde{c}_{n+1}  
         \nonumber \\ 
         &&  ~  +  
          \sqrt{2 ( 1 - \epsilon_{n} ) } \, c_{n} s_{n} \, \tilde{s}_{n+1}  ~ ] \, \tilde{s}_{n+2} ~ \big| ^2 ~ ,
\end{eqnarray}
as well as
\begin{eqnarray} \label{u2_3}
         {\cal U}_{bb , n} &=&
           \big| ~  [ ~ 2 \sqrt{ \epsilon_{n} ( 1 - \epsilon_{n} ) } s_{n+1}^2 \, \tilde{c}_{n+1} \, 
             \nonumber\\ 
             && ~~\,  +    
             \sqrt{2 ( 1 - \epsilon_{n} ) } \,  c_{n-1} s_{n+1} \, \tilde{s}_{n+1} ~ ] \, \tilde{c}_{n+2} 
            +  
             \nonumber \\ 
             && 
             ~~ [  \,   c_{n}^2 \, \tilde{s}_{n+1} 
             \, + \, 
             \sqrt{2 \epsilon_{n} }  c_{n} s_{n} \, \tilde{c}_{n+1} \, ]  \, \tilde{s}_{n+2}  \,  |^2  ~ .
\end{eqnarray}

   \noindent
In these expression we make use of the convenient  definitions
\begin{eqnarray} 
  \epsilon_n &\equiv& \frac{n+1}{2n+3} ~ ,
  \\ \nonumber
\end{eqnarray} 
and trigonometric functions for the one-atom events
\begin{eqnarray} 
  \tilde{s}_n &=&  \sin[ \, g s \sqrt{n} \, ]  ~ , 
  \\
  \tilde{c}_n &=&  \cos[ \, g s \, \sqrt{n} \, ] ~ ,
  \end{eqnarray} 
as well as for the two-atom events, i.e.
\begin{eqnarray}
  s_n &=& \sin[ \, g (\tau-s) \, \sqrt{n+3/2} \, ] ~ ,
  \\
  c_n &=& \cos[ \, g (\tau-s) \, \sqrt{n+3/2} \, ] ~ .
\end{eqnarray}

   \noindent
   The expressions we obtain are the same as in Ref.\cite{schieve_03}. We also mention that,
   due to unitarity, the two-atom matrix satisfies
\begin{eqnarray} 
        \sum_{n=0}^{\infty} ~ ({\cal U}_2)_{nm} &=& 1 ~ ,
\end{eqnarray}

     \noindent
     for any given integer $m \geq 0$. This can be verified
     analytically. It is also now straightforward to verify the
     relation ${\cal U}_1 ^2 = {\cal U}_2(\tau)$, which corresponds to the case when the atoms in the beam
     are separated in time by $s=\tau$, i.e. no overlap.

     In particular, let us consider the limit in which the mean number of atoms in the cavity
     is small, i.e. $\epsilon \ll 1$. In this case the micromaser is close to the one-atom maser
     situation. From Eqs. (\ref{P_1_t}) and (\ref{P_2_t}) we then obtain $\tilde{P}_1 \approx 1 - 2 \epsilon$
     and $\tilde{P}_2 \approx \epsilon$. Furthermore, the conditional Poissonian probability distribution 
Eq.(\ref{f_p}) 
     is then reduced to $w(s) \approx 1/\tau$. We then obtain

\begin{eqnarray}     \label{master_2_atom_limit}
   \frac{d p }{d t} &\approx& - \gamma \, \bigg \{ \, L_C  -  N \, (
   \, 1 - 2 \epsilon \, )  \, ( \, {\cal U}_1 - 1  \, )  
                    \\ \nonumber
                    && ~~~~~~~ -   N \, \epsilon \, ( \, {\cal U}_2 - 1  \, ) \, \bigg \} \, p ~ ,
\end{eqnarray}

     \noindent
     where

\begin{eqnarray}
  ({\cal U}_2)_{nm}  &\approx& \frac{1}{\tau} \, \int_0^{\tau} ds ~ ({\cal U}_2 (s))_{nm} ~ .
\end{eqnarray}

     \noindent
 The master equation Eq.(\ref{master_2_atom_limit}) appears to be of the same form as  the result in 
Ref.\cite{haake97} (see also Refs.\cite{ulzega_99,skagerstam_06}). There is, however, one important difference.
In Ref.\cite{haake97} it is assumed that both $\epsilon$ and the one-atom generator ${\cal U}_1 -1$ are to be treated as small entities. This e.g.  means that ${\cal U}_2(\tau)= 1+ 2({\cal U}_1-1)+{\cal O}(({\cal U}_1-1)^2)$. In order to compare Eq.(\ref{master_2_atom_limit}) with the corresponding expression of Ref.\cite{haake97}, we should therefore in general only retain terms to at most second order in the one-atom trigonometric functions  $\tilde{c}_n$ and $\tilde{s}_n$  in Eqs.(\ref{u2_1})-(\ref{u2_3}). With these approximations we arrive at the same expression as in Ref.\cite{haake97} as is also used in Ref.\cite{skagerstam_06}. In our numerical simulations to be presented below we make use of the exact expression for ${\cal U}_2$.

\begin{figure}[ht]

\begin{picture}(0,0)(122,660)   


\includegraphics{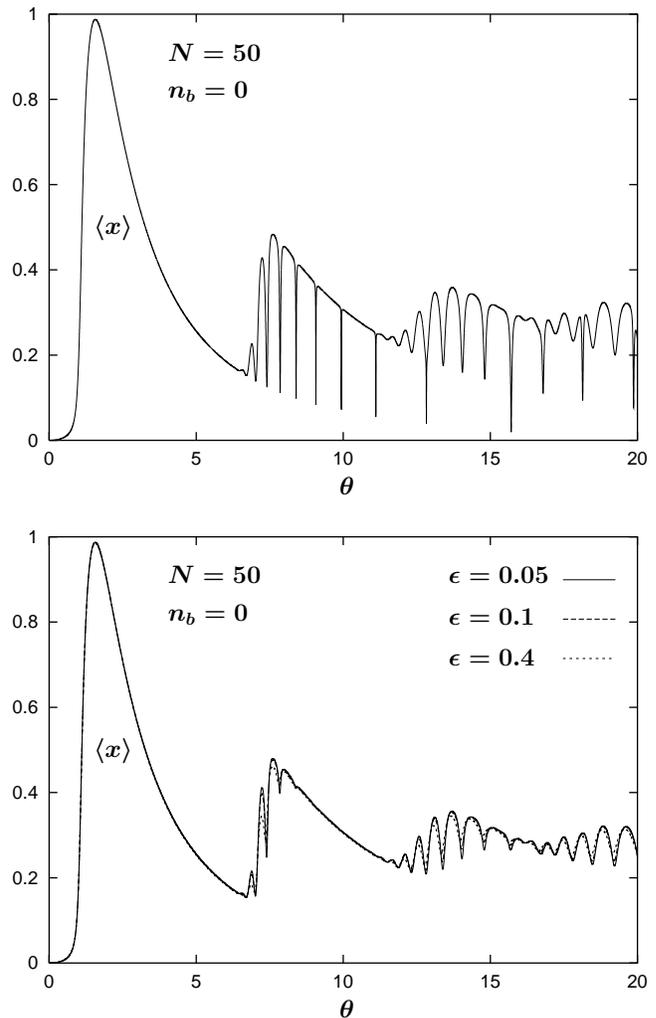}

\end{picture}

\vspace{13.5cm}

\caption{The order parameter $\langle x \rangle = \langle n/N \rangle$
         as a function of $\theta$ for $N=50$, $n_b=0$. The upper graph shows
         the order parameter for the one-atom micromaser, i.e. $\epsilon = 0$.
         The lower graph shows the micromaser when two-atom events are included with
         $\epsilon = 0.05 , 0.1 , 0.4$.
        }
\label{Kolobov_figs_1}

\end{figure}

    Our final expression for ${\cal L}_2$ as given in Eq.(\ref{eq:cns}) is the same as in Ref.\cite{schieve_03}.
    It is rather straightforward to implement the matrix elements of ${\cal L}_2$ in a numerical routine
    in order to find the stationary probability eigenvector ${\bar p}$ corresponding to the eigenvalue 
    $\lambda_0=0$ of ${\cal L}_2$. This eigenvector ${\bar p}$ is to be used in evaluating the mean number of photons $\langle n \rangle$
    in the cavity. 
    The next-to-lowest eigenvalue $\lambda_1$ of ${\cal L}_2$, which we find numerically, will then determine 
    typical time scale  for the approach to the stationary situation as described by the eigenvector ${\bar p}$. The joint probability for observing
    two atoms, with a time-delay $t$ between them, can now be used in order to define a correlation length   $\gamma^A(t)$ as first suggested by Lautrup {\it et al.} 
  \cite{ElmforsLS95} (for a recent discussion see Ref.\cite{skagerstam_06}). At large times $t\rightarrow \infty$, 
  the atomic beam correlation length is then given by $\xi_A$ by \cite{ElmforsLS95}, i.e.
\begin{equation}
   \gamma_A(t) \sim e^{-t/\xi_A} ~,
\end{equation}
 \noindent
   which then  is determined by $\lambda_1$, i.e. $\gamma\xi_A=1/\lambda_1$.
   In passing we notice that for photons a similar correlation length $\xi_{C}$ was also defined in Ref.\cite{ElmforsLS95} in terms of correlations of the number of photons. It
   was shown that, quit naturally,  these correlation lengths are identical, i.e. $\xi_A=\xi_C\equiv \xi$ \cite{ElmforsLS95}.
   In the numerical work presented below, it turns out to be sufficient to use $200$x$200$ matrices for ${\cal L}_2$
   in order to obtain the accuracy of the graphs as presented in the present paper. 

\begin{figure}[ht]

\begin{picture}(0,0)(122,660)   

\includegraphics{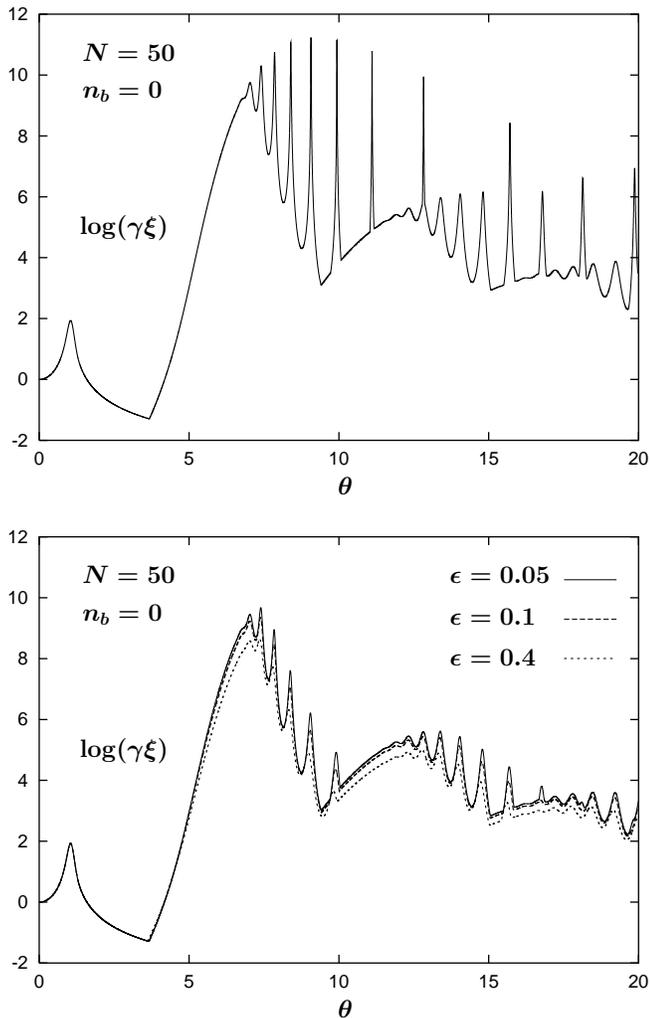}

\end{picture}

\vspace{13.5cm}

\caption{The correlation length $\gamma\xi$ as a function of $\theta$ for $N=50$, $n_b=0$. The upper graph shows
         the correlation length for the one-atom micromaser, i.e. $\epsilon = 0$.
         The lower graph shows the micromaser when two-atom events are included with
         $\epsilon = 0.05 , 0.1 , 0.4$.
        }
\label{Kolobov_figs_2}

\end{figure}

\vspace{0.5cm}
\bc{
\section{RESULTS}
\label{sec:results}}
\ec
\vspace{-0.5cm}
It was argued in Refs.\cite{haake97,skagerstam_06} and in Ref.\cite{casagrande99b} that trapping
 effects,
 which occur at $\theta =k\pi \sqrt{N/n}$, 
       ($k,n=1,2,...$), may be significantly suppressed due to the inclusion of collective two-atom events.
       For reasons of comparison, Fig.\ref{Kolobov_figs_1} and Fig.\ref{Kolobov_figs_2} in the present paper are plotted 
       for the same parameters as in Refs.\cite{haake97,skagerstam_06}.
       In Fig.\ref{Kolobov_figs_1} we show the results of a numerical evaluation of the order parameter $\langle x\rangle$, where $x=n/N$ and $\langle ~\rangle$ 
       as above denotes an average with respect to the stationary solution ${\bar p}$ when two-atom collective effects are included. For $\epsilon = 0$, the non-trivial and detailed structure of the various trapping signals, i.e. the observed dips in
$\langle x\rangle$,   has been explained in Ref.\cite{Rekdal&Skagerstam&99a}. The numerical precision of our calculations, as well as in Ref.\cite{skagerstam_06},  is sufficient to reveal the exact large-$N$ structure of these dips in the order parameter $\langle x\rangle$. This is  not the case in Ref.\cite{haake97}. 
        We, furthermore, observe a discrepancy between our numerical results and the corresponding
       results in Refs.\cite{haake97,skagerstam_06}. This discrepancy is mainly due to fact that we make use of the exact form of the two-atom generator ${\cal U}_2$ which was not the case in Refs.\cite{haake97,skagerstam_06}. In fact, it turns out that the exact expression for the conditional probability density Eq.(\ref{f_p}) and the modified probabilities Eq.(\ref{P_1_t}) and Eq.(\ref{P_2_t}) are of less importance.
We have verified this by reconsidering the numerical evaluations of Ref.\cite{skagerstam_06} in terms of the Eqs.(\ref{f_p}), (\ref{P_1_t}) and (\ref{P_2_t})instead of using their small $\epsilon$-expansions, and we find only small numerical corrections to results presented in Ref.\cite{skagerstam_06}.
An exact treatment of the two-atom generator ${\cal U}_2$ therefore, remarkably, leads to smaller collective effects as compared to the results of Refs.\cite{haake97,skagerstam_06}. Most of the dips in $\langle x\rangle$ corresponding to  trapping states actually do not vanish for $\epsilon \neq 0$ in our case, even for mean number of atoms inside
       the cavity as large as $\epsilon = 0.4$. This conclusion is in agreement with the findings of Ref.\cite{schieve_03}.
 
\begin{figure}[ht]
    \begin{picture}(0,0)(122,665)   

\includegraphics{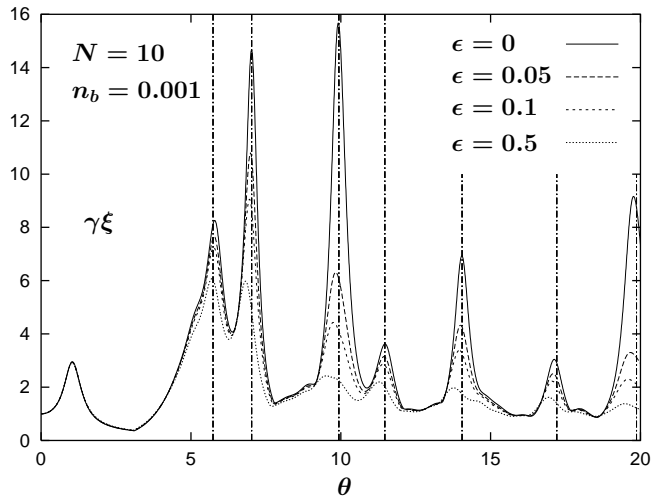}

\end{picture}

\vspace{6.6cm}

\caption{  The correlation length  $\gamma \xi$ as a function of 
           $\theta$ for various values of $\epsilon$.
           The vertical lines indicate the trappings values at 
           $\theta= \pi \sqrt{N}(1/\sqrt{3}, 1/\sqrt{2},1,2/\sqrt{3},\sqrt{2}, \sqrt{3},2) $.  }
\label{corr_fig_001}
\end{figure}

 As mentioned above, it may, however, be difficult to experimentally discriminate between the collective effects and other corrections, like the effects of detection efficiencies, when only considering the order parameter $\langle x\rangle$. As seen from Fig.\ref{Kolobov_figs_2}, where the correlation  length $\gamma \xi$is plotted on a logarithmic scale,
trapping states are more pronounced in terms of this observable as compared to the order parameter $\langle x\rangle$ as given in Fig.\ref{Kolobov_figs_1}. Even for such a low value
       of the atomic flux parameter as  $N=10$, and with $n_b=0.001$ as in Fig.\ref{corr_fig_001},
the trapping states are clearly revealed.
       These results therefor convincingly shows that trapping states can be even more clearly revealed experimentally in terms of the correlation length $\gamma \xi$ as compared to the order parameter $\langle x \rangle$.  In this context it is important to notice that the effect of detection efficiencies on the correlation length $\gamma \xi$ easily can be taken into account  in terms of a simple renormalization of the experimental data as discussed in detail in Ref.\cite{skagerstam_06}.   

\begin{figure}[ht]

\begin{picture}(0,0)(122,665)   

\includegraphics{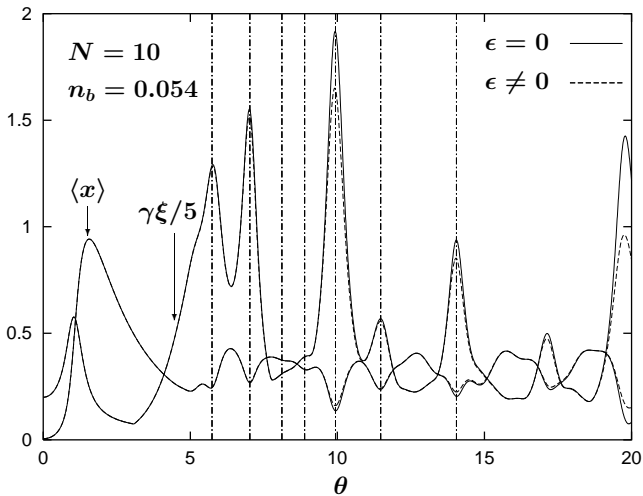}

\end{picture}

\vspace{6.6cm}

\caption{  The order parameter $\langle x \rangle$ and the correlation length  $\gamma \xi/5$ as a function of 
           $\theta$ with no collective effects taken into account 
($\epsilon =0$) or $\epsilon =R\tau = \theta\sqrt{N}\gamma/g$.
           The Rabi frequency is as in the experimental study of trapping-states 
in Ref.\cite{W&V&H&W}, i.e. $g=39\,$kHz and $\gamma = 10~s^{-1}$.
           The vertical lines indicate the experimentally observed trappings values at 
           $\theta =\pi \sqrt{N}(1/\sqrt{3}, 1/\sqrt{2},2/\sqrt{6},2/\sqrt{5},1,2/\sqrt{3},\sqrt{2}) $ as reported in Ref.\cite{W&V&H&W}. }
\label{garching_trapping_99}
\end{figure}

       In the experimental study of trapping states in Ref.\cite{W&V&H&W,W&V&H&W_2} the parameters are 
       varied in such a way that the parameter $\epsilon$ is not constant but is given by 
       $\epsilon \equiv R\tau = \theta\sqrt{N}\gamma/g$. In Fig.\ref{garching_trapping_99}, we have chosen 
       the cavity temperature to be $T=0.3$ K, corresponding to $n_b = 0.054$, and a low pump rate
       corresponding to $N=10$. 
       These particular choices of parameters are the same as in Weidinger {\sl et al.}~\cite{W&V&H&W}, in which case two-atom events are rare, i.e. $\epsilon \leq \epsilon(\theta=20) = 0.016 \ll 1$.
       The solid line in Fig.\ref{garching_trapping_99} corresponds to $\epsilon=0$, i.e. no 
       collective effects taken into account. 
       Comparison with the dashed line, where two-atom collective effects are taken into account, gives basically
       the same result. This is not unreasonable as the mean number of photons inside the cavity is very small in
       this case. We also mention that, for the particular parameters under consideration,
       the result in Ref.\cite{skagerstam_06} (see their Fig. 6) gives basically the same result.

\begin{figure}[t]

\begin{picture}(0,0)(115,665)   

\includegraphics{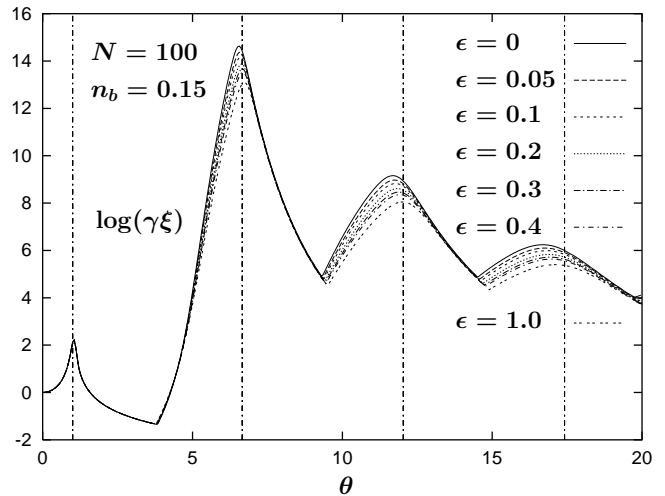}

\end{picture}

\vspace{6.6cm}

\caption{  The correlation length  $\gamma \xi$ as a function of $\theta$ for various values of $\epsilon$. 
           The vertical lines indicate the $\theta$ values of the large $N$ one-atom maser phase transitions at 
           $\theta^{*}_{0}=1 , \theta^{*}_{01}\approx 6.6610, \theta^{*}_{12}\approx 12.035, \theta^{*}_{23}\approx 17.413 $. 
           With $\langle x\rangle$ as an order parameter, the transition at $\theta= \theta^{*}_{0}$
           is second-order while the other are first-order transitions \cite{Rekdal&Skagerstam&99a}. }
\label{Phase_Transitions}

\end{figure}

     In Refs.\cite{ElmforsLS95,Rekdal&Skagerstam&99a} the phase structure of the micromaser 
     system in terms of the physical parameters at hand is investigated.
     It is shown that there are two types of transitions, a thermal-to-maser transition
     and maser-to-maser transitions. In the large $N$ limit, the thermal-to-maser transition occurs 
     at $\theta_0^*$ and the maser-to-maser transitions occur at $\theta_{kk+1}^*$ for $k \geq 0$
     in the absence of collective effects.
     For given values of the parameters, these values of the pump parameter are computed numerically.
     At finite $N$ and including collective pump atom effects, we can see from e.g.
     Fig.\ref{Kolobov_figs_2} and in particular Fig.\ref{Phase_Transitions} that signals of the large
     $N$ phase transitions in the correlation length
     $\gamma \xi$ are still clearly exhibited close to the corresponding one-atom maser phase transitions.
     As pointed out in Refs.\cite{ElmforsLS95,Rekdal&Skagerstam&99a,skagerstam_06}, the critical point $\theta^{*}_{0}$
     of the first second-order maser transition is clearly exhibited in both  $\langle x \rangle$ as well as
     $\log( \gamma \xi )$. This fact is illustrated for $\gamma \xi$ in Fig.\ref{Phase_Transitions}.
     Furthermore, in Ref.\cite{skagerstam_06} using a perturbative reasoning, it is argued that, for sufficiently small $\epsilon$,
     the critical parameters $\theta^{*}_{k+1} \propto \sqrt{N}$ of first-order transitions
     are changed to ${\bar\theta}^{*}_{kk+1}\equiv \exp(-\epsilon)\theta^{*}_{kk+1}$ 
     and that $\log(\gamma\xi)_{crit}$ is changed to 
     $\log(\gamma{\bar \xi})_{crit} \equiv \exp(-\epsilon)\log(\gamma\xi(\theta = {\bar\theta}^{*}_{kk+1}))$. This change of the correlation length was also in agreement with the numerical result of Ref.\cite{skagerstam_06}.
     When using the exact expression for the two-atom generator ${\cal U}_2$, these perturbative considerations are, however, not valid as e.g. can be seen from our numerical results presented in
     Fig.\ref{Phase_Transitions}. In this figure we see that 
     the peak values of $\log(\gamma{\xi})_{crit}$ are actually shifted slightly to higher rather then to smaller values of
     the pump parameter $\theta$. In Fig.\ref{Phase_Transitions} we have also evaluated the correlation length for $\epsilon = 1$ and we still find that collective two-atom effects are small.

\vspace{0.5cm}
\bc{
\section{CONCLUSIONS}
\label{sec:final}
}\ec

%
%

     In conclusion, we have studied two-atom collective effects in the micromaser system. 
     We have confirmed the result of Ref.\cite{schieve_03} that the characteristic effects of trapping states on the order parameter are not as suppressed as was previously 
suggested \cite{haake97,skagerstam_06}. We have seen that the major reason for this is connected to the exact treatment of the two-atom generator. More importantly, 
 the presence of trapping states are more pronounced in terms of the correlation length $\gamma \xi$ 
     as compared the order parameter $\langle x\rangle$. This conclusion remains true when collective two-atom effects are taken into account even for a large average value of atoms inside the cavity.
As seen in  Fig.\ref{garching_trapping_99}, which refers to an actual experimental situation \cite{W&V&H&W},  the vacuum trapping state, which for the actual micromaser parameters occurs at $\theta =\pi\sqrt{N} \approx 9.94$, is clearly more visible in the correlation length than in the average number of photons. This conclusion also remains true when collective two-atom effects are taken into account. An experimental preparation of the  vacuum trapping state is essential when considering axion detection in the micromaser system \cite{jones_2006}. We conclude that trapping states can be more clearly revealed experimentally in terms of the atomic correlation length than in terms of the natural micromaser order parameter $\langle n/N \rangle$. This observation is apparently in accordance with the recent analysis of Ref.\cite{jones_2006}.

\vspace{1cm}

{\begin{center}
{ \bf ACKNOWLEDGMENT }
\end{center}}
%

      One of the authors (P.K.R.) wishes to thank U. Hohenester for discussions and warm hospitality
      while parts of the present work was completed. The research has been supported in part by the Austrian 
      Science Fund (FWF). The research of B.-S.S. was supported by the Norwegian University of Science and Technology.

\end{document}